# Unsupervised Co-segmentation of 3D Shapes via Functional Maps

**Jun Yang**

*School of Electronic and Information Engineering, Lanzhou Jiaotong University, Lanzhou 730070, P. R. China*

*yangj@mail.lzjtu.cn*

**Zhenhua Tian**

*School of Automation and Electrical Engineering, Lanzhou Jiaotong University, Lanzhou 730070, P. R. China*

*786578144@qq.com*

**Abstract**：We present an unsupervised method for co-segmentation of a set of 3D shapes from the same class with the aim of segmenting the input shapes into consistent semantic parts and establishing their correspondence across the set. Starting from meaningful pre-segmentation of all given shapes individually, we construct the correspondence between same candidate parts and obtain the labels via functional maps. And then, we use these labels to mark the input shapes and obtain results of co-segmentation. The core of our algorithm is to seek for an optimal correspondence between semantically similar parts through functional maps and mark such shape parts. Experimental results on the benchmark datasets show the efficiency of this method and comparable accuracy to the state-of-the-art algorithms.

**Keywords:** Functional Maps, Co-segmentation, K-means Clustering, Mesh Laplace Operator, Heat-kernel Signature

## 1 Introduction

In recent years, there has been a highly interest in high-level co-analysis of sets of 3D shapes. Current research work has demonstrated that more semantic knowledge can be extracted by simultaneously analyzing a set of shapes, rather than analyzing each shape individually, and better segmentation results can be obtained by applying this semantic knowledge on co-segmentation of a set of shapes [1-4].

Co-segmentation is a fundamental task in high-level co-analysis of shapes, which has witnessed increasing interests. It simultaneously segments a set of shapes from the same class into consistent semantic parts with correspondence [5-10]. Consistent segmentation has been utilized in vast areas including modeling [11], shape retrieval [12], texturing [13], etc.

However, extraction of appropriate knowledge inherent to multiple shapes for consistent segmentation remains challenging [14].

In this paper, a novel framework for unsupervised co-segmentation of a set of 3D shapes from the same class is proposed. We firstly segment all input shapes into the primitive meaningful parts. Then the correspondence between parts with same semantic information is constructed, which is then labeled indicating a classification. Finally, results of co-segmentation are obtained by labelling the various parts of the input shapes by classification originally established. These shapes loosely belong to the same class, implying a consistency in the composition of their major functional components. However, the shapes may exhibit a high degree of geometric variability as well as variation in their finer-scale structures as shown in Figure 1. The key point of this method is to build links between the same semantic parts, which ensure consistence of co-segmentation.

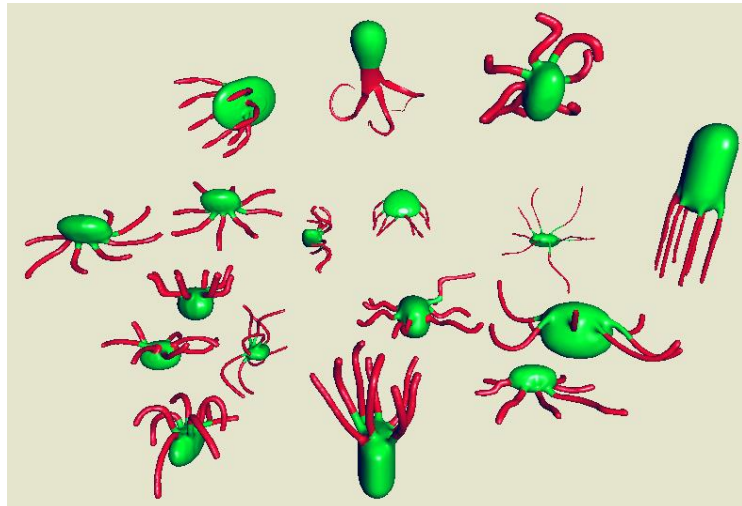

Figure 1. Unsupervised co-segmentation results of our approach. Corresponding parts with highly dissimilarities in their geometry and topology are shown with same color

## 2 Related work

Shape segmentation is a fundamental task in shape analysis, which decomposes a 3D shape into meaningful parts. A large variety of approaches have been proposed for segmenting single shape into meaningful parts. It has been shown that no segmentation algorithm always performs well for all models because the geometry of an individual shape may lack sufficient cues to identify all parts that would be perceived as meaningful to a human observer [15]. So some researchers focus on consistently segmenting a set of shapes from same class into meaningful parts, then compared to classical shape segmentation, which not only segments shapes, but also establishes correspondence between them[16, 17].

Golovinskiy et al. [1] simultaneously segment models and create correspondences between segments. A graph is constructed whose nodes represent the faces of every mesh, and whose edges connect adjacent faces within a mesh and corresponding faces in different meshes. Then a consistent segmentation is created by clustering this graph. To deal with non-homogeneous part scales, Xu et al. [2] factor out the scale variation in the shape parts by first classifying the shapes into different styles. In this manner, they are able to co-segment shapes with a higher level of variability. However, the graph generation process is computationally expensive.

Kalogerakis et al. [5] propose a data-driven approach for simultaneous segmentation and labeling of parts in 3D meshes. An objective function is formulated as a conditional random field model, with terms assessing the consistency of faces with labels, and terms between labels of neighboring faces. The objective function is learned from a collection of labeled training meshes. Van Kaick et al. [14] introduce an approach to part correspondence which incorporates prior knowledge imparted by a training set of pre-segmented, labeled models and combines the knowledge with content-driven analysis based on geometric similarity between the matched shapes. While the idea of joint labeling is quite general, the content analysis component of their approach is still fairly primitive in terms of the feature similarity employed.

Recently, some promising unsupervised techniques have been proposed for co-segmentation. Huang et al. [4] present an entirely unsupervised joint segmentation approach based on an integer quadratic programming formulation solved via a linear programming relaxation using a block coordinate descent procedure that makes the optimization feasible for large databases. They evaluate their approach on the Princeton segmentation benchmark and show comparable results to the supervised approaches. The limitation of this approach is that final segmentation depends on initially computed patches. Sidi et al. [6] pose the co-segmentation problem as that of clustering in a descriptor space, which allows their method to handle shapes with rich variations in part composition and geometry, where a rigid alignment scheme would not lead to a proper co-segmentation.

Hu et al. [7] deal with the co-segmentation as a subspace clustering problem in multiple features spaces, which generates the segmentations by clustering the primitive patches of the shapes in subspace and obtains their correspondences simultaneously. This technique does not guarantee that all co-segmentations can be properly captured, since the final segmentations of each shape absolutely rely on the initially computed patches. Subsequently, Wu et al. [18] improve this method through fusing multiple descriptors and integrating the feature selection and spectral clustering into a unified procedure. Nevertheless, this will not work if all descriptors cannot properly characterize the shape. Very recently, Meng et al. [9] build a statistical model to describe cluster of each parts, and employ the multi-label optimization scheme to refine final co-segmentation of input meshes. But it may fail if the corresponding parts are very difficult to classify.

## 3 Co-segmentation

Given a set of 3D shapes from the same class, the goal of our algorithm is to obtain segmentation and labeling of the shapes that is consistent across the set. The input shapes have a similar semantic part composition and thus share a common label set that reveals correspondence between shapes. Our approach consists of three steps to carry out the co-segmentation. First, each shape is decomposed into primitive patches independently. Then, labels representing a class of segments are obtained according to correspondences between candidate parts. Finally, co-segmentation is achieved by labeling the shapes based on the calculated correspondence.

### 3.1 Pre-segmentation

The first step in our algorithm is to segment every shape of the input set into smaller candidate parts individually. We firstly compute descriptors on 3D models according to mesh Laplace operator, and then obtain a set of candidate segments by using K-means algorithm to cluster patches into meaningful semantic parts. What we focus on is to analyze part of shapes, rather than primary patches.

The feature descriptor matrix $W$ for each model is calculated based on mesh Laplace operator. Suppose that we have constructed a $k$-dimensional embedded space by the first $k$ maximum eigenvalues $\lambda_1, \lambda_2, \cdots, \lambda_k$ and the corresponding eigenvectors $\phi_1, \phi_2, \cdots, \phi_k$ of $W$. The coordinates of vertices of 3D meshes in the embedded space are represented as following.

$$Co = \left[ \frac{\phi_1}{\sqrt{\lambda_1}}, \frac{\phi_2}{\sqrt{\lambda_2}} \cdots, \frac{\phi_k}{\sqrt{\lambda_k}} \right] \quad (1)$$

Similarities between shapes can be described easily in Euclidean distance among different vertices in the embedded space.

The initial pre-segmentation for each model is calculated by combining the $k$-means clustering algorithm with coordinates $Co$. We only need to set a parameter $n$, i.e., quantity of segmentation parts, during the process of clustering which is based on Euclidean distance from cluster center to vertex. The final results of pre-segmentation are achieved till the iterative refinement procedure of clustering converges.

### 3.2 Functional maps

We employ Functional maps [19] to build correspondence of parts which have the same semanteme. Because the same semantic parts of different shapes usually have diversities of geometric structures, geometric feature descriptors are not suitable for reflecting an inherent relationship between them, whereas functional maps representation is natural for clustering according to their intrinsic relationships. Therefore, we can create shape correspondences between meaningful semantic parts by using functional maps, resulting in good co-segmentation of sets of shapes.

Given a pair of shapes $M$ and $N$, correspondence between them can be represented as a mapping problem $T: M \to N$, $T$ induces a natural transformation of derived quantities, such as functions on $M$. To be precise, if we define a scalar function $f: M \to \mathbb{R}$, then we obtain a corresponding function $g: N \to \mathbb{R}$ by composition, as in $g = f \circ T^{-1}$. Therefore, it can denote this induced transformation by $T_F: F(M, \mathbb{R}) \to F(N, \mathbb{R})$, where $F(\bullet, \mathbb{R})$ denotes a generic space of real-valued functions. $T_F$ can be called the functional representation of the mapping $T$.

Now suppose that the function space of $M$ is equipped with a basis so that any function $f: M \to \mathbb{R}$ defined on the shape can be represented as a linear combination of basis functions $f = \sum_i a_i \phi_i^M$, where $a_i$ is a vector of coefficient, and $\phi_i^M$ is the $i^{\text{th}}$ eigenvector of mesh Laplace on shape $M$. Then, mapping of function $f$ can be represented as

$$T_F(f) = T_F\left(\sum_i a_i \phi_i^M\right) = \sum_i a_i T_F\left(\phi_i^M\right) \tag{2}$$

In addition, If $N$ is equipped with a set of basis function $\{\phi_j^N\}$, then $T_F\left(\phi_i^M\right) = \sum_j c_{ij} \phi_j^N$ is established for some $\{c_{ij}\}$ and

$$T_F(f) = \sum_i a_i \sum_j c_{ij} \phi_j^N = \sum_j \sum_i a_i c_{ij} \phi_j^N \tag{3}$$

Let $\{\phi_i^M\}$ and $\{\phi_j^N\}$ be bases for $\mathcal{F}(M, \mathbb{R})$ and $\mathcal{F}(N, \mathbb{R})$ respectively. A generalized linear functional mapping $T_F: \mathcal{F}(M, \mathbb{R}) \to \mathcal{F}(N, \mathbb{R})$ with respect to these bases is the operator defined by

$$T_F\left(\sum_i a_i \phi_i^M\right) = \sum_j \sum_i a_i c_{ij} \phi_j^N \tag{4}$$

where $c_{ij}$ is the $j^{\text{th}}$ coefficient of $T_F\left(\phi_i^M\right)$ in the basis $\{\phi_j^N\}$. The mapping $T_F$ can be represented via computing a matrix $\boldsymbol{C}$. Note that $\boldsymbol{C}$ has a particularly simple representation if the basis functions $\{\phi_i^N\}$ are orthonormal with respect to some inner product $\langle \cdot, \cdot \rangle$, namely $c_{ij} = \langle T_F\left(\phi_i^M\right), \phi_j^N \rangle$. So we conclude for any function $f$ represented as a vector of coefficients $\boldsymbol{a}$ then

$$T_F(\boldsymbol{a}) = \boldsymbol{Ca} \tag{5}$$

### 3.3 Mesh Laplace

We used the mesh Laplace eigenfunctions as the basis for the function space on each shape,

which provides a natural multi-scale way to approximate functions. The mesh Laplace is the first algorithm with point-wise convergence that is guaranteed for arbitrary meshes. Let $U$ be a mesh in $\mathbb{R}^3$ and $V$ be a set of vertices of the mesh $U$. This algorithm takes $f:V \to \mathbb{R}$ as input and $L_U^h f:V \to \mathbb{R}$ as output. The discrete mesh Laplace operator is defined, for any vertex $\omega \in V$, as follows [20]:

$$L_U^h f(\omega) = \frac{1}{4\pi h^2} \sum_{X \in U} \frac{Area(X)}{\#X} \times \sum_{p \in V(X)} e^{-\frac{\|p-\omega\|}{4h}} \left( f(p) - f(\omega) \right) \tag{6}$$

For a face $X \in U$, $Area(X)$ denotes the area of $X$, the number of vertices in $X$ is denoted by $\#X$, and $V(X)$ is the set of vertices of $X$. The parameter $h$ is a positive quantity, which intuitively corresponds to the size of the neighborhood considered at each point.

To solve this problem in matrix form, we define weighted adjacency matrix $\mathbf{W}=(w_{ij})$, diagonal matrix $\mathbf{Q}=diag(q_1,\cdots,q_n)$, the lumped mass matrix $\mathbf{D}=(d_1,\cdots,d_n)$, stiffness matrices $\mathbf{A}=\mathbf{Q}-\mathbf{W}$. Then the mesh Laplace matrix can be defined as $\mathbf{L}=\mathbf{D}^{-1}\mathbf{A}$. Finally, the equation (6) can be written as a generalized symmetric problem $\mathbf{A}f = \lambda \mathbf{D}f$. The eigenvalues and eigenvectors can be obtained and be defined as $\overline{\lambda}=(\lambda_1,\cdots\lambda_i,\cdots\lambda_n)$ and $\overline{\boldsymbol{\Phi}}=(\phi_1,\cdots\phi_i,\cdots\phi_n)$ respectively. We use the first $n$ (here $n$=50) mesh Laplace eigenfunctions as the basis for their functional representations. Since eigenfunctions of the mesh Laplace operator are ordered from "low frequency" to "higher frequency," meaning that they provide a natural multi-scale way to approximate functional mappings between shapes. Even if a 3D model undergoes near-isometric deformation, the stable functional space can still be obtained.

3.4 Co-segmentation

Through the functional maps, correspondence between shape components is obtained, and co-segmentation signature is used to represent the correspondence. We mark input shapes in model space to obtain the combination of the segmentation results of a set of 3D shapes.

Firstly, each shape is pre-partitioned resulting in significant difference of the same semantic components in the descriptor space, due to lack of intrinsic semantic relation between the segmented parts, as shown in Figure 2. The noses of pliers in Figure 2(a) have the same semantic information, however, their geometric characteristics are different, this leads to significant overlap of the nose and handle of the plier in the descriptor space, also the semantic difference between the cutter and the other parts of the plier is not distinct. We use the functional maps to establish correspondence between the segmented parts and linked together their semantic parts through the intrinsic relationship between them, this leads to a more semantically consistent representation of each parts of the plier in the descriptor space.

Therefore, the same parts are clustered together, and each cluster is a signature of the parts in the co-segmentation of the descriptor space. Secondly, we use each signature which carries a label to represent the original model space, and therefore a final co-segmentation result is obtained, as shown in Figure 2(b).

# 4 Calculation of correspondence

In this section we mainly discuss implementation of the co-segmentation process by using functional map to establish and obtain correspondence between same semantic segments of sets of 3D shapes, then introduce the choice of basis space and function constraints.

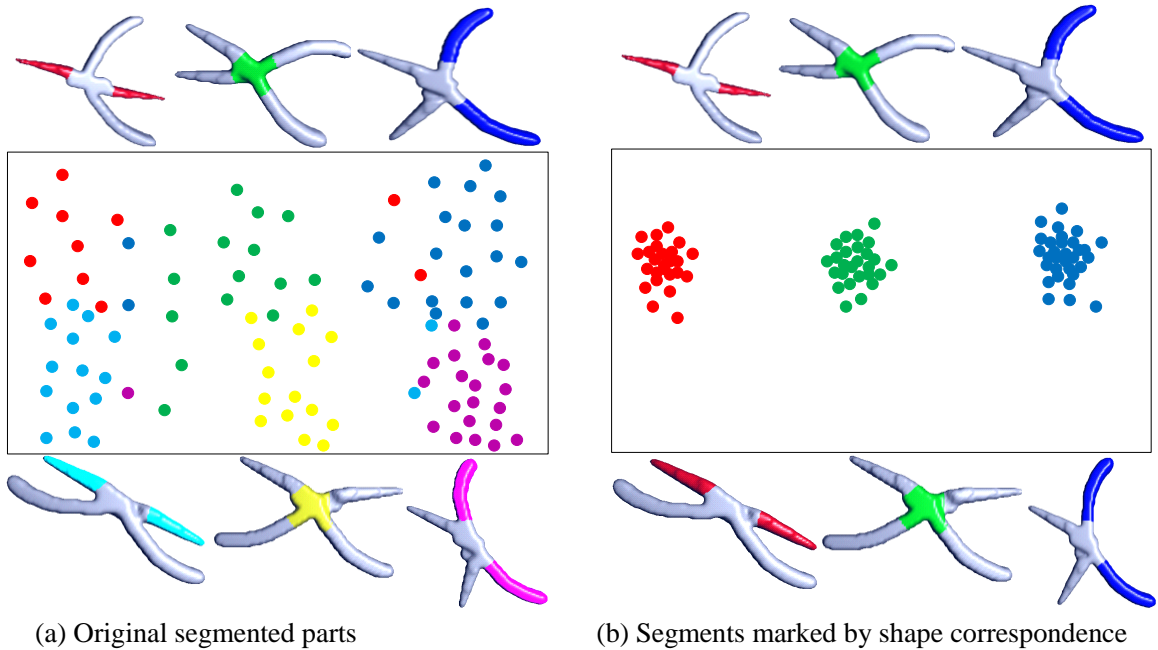

(a) Original segmented parts        (b) Segments marked by shape correspondence

Figure 2 Marks of co-segmentation

## 4.1 Choice of function for basis space

Compactness and stability are the two most important features for choosing function of basis space. Compactness means that most intrinsic functions on a shape should be well approximated by using a small number of basis elements, while stability means that the space of functions spanned by all linear combinations of basis functions must be stable under small shape deformation. These two characteristics together make it possible to express the mapping $T_F$ using a small, more efficient subset of the basis functions. As a result, we use the eigenfunctions of mesh Laplace operator as a basic space of model.

Given two shapes $M$ and $N$, we respectively used the first $k$ eigenfunctions $\phi_i^M$ and $\phi_j^N$ of mesh Laplace operator as the bases of function space. A mesh Laplace operator is a local descriptor, where different feature information is represented by different feature function. Here, "higher frequency" feature function contains less information. Since eigenfunctions are ordered from "low frequency" to "higher frequency", a few of the low frequency feature functions can be well expressed as characteristic of shape. If several eigenfunctions of the mesh Laplace operator of shape are selected, then we can obtain a multi-scale space base for the shape. The distribution of different eigenvector of mesh Laplace on the shape is shown in Figure 3. From left to right, $n$, the order of eigenvector, is the 2nd, 8th, 10th, 15th and 20th.

Using the mesh Laplace operator as a basis space for the shape, not only provides a multi-scale feature, but also ensures that the calculations of shape correspondence are sparse and therefore can be effectively stored. When we calculate the correspondence between shapes by the functional maps, if the first 50 eigenfunctions are selected as a basis space for the shapes, then the matrix $C$ will be of size 50x50, however, if not, instead using $10k$ point clouds of a shape will require a matrix of size $10k$ x $10k$ to represent such a correspondence. Therefore, we can intuitively see that storage is preserved using functional map to find correspondence between shapes.

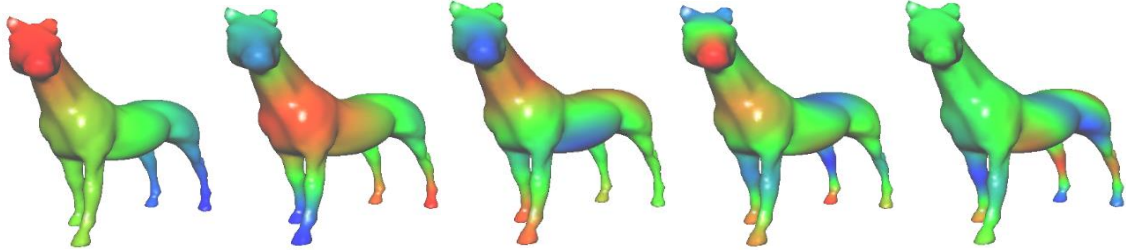

Figure 3 Distribution of different eigenvector of mesh Laplace on the shape

In theory, if the shapes $M$ and $N$ are isometric, then the matrix $C$ of functional map will also have a constant volume. However, in practice, shapes $M$ and $N$ are approximately isometric, and as such the matrix $C$ is approximately sparse or is in the shape of a funnel. Additionally, in general we obtain the matrix $C$ such that at least 90% of the values of its elements are less than 0.1, this guarantees a sparse matrix since these values can often be ignored.

### 4.2 Descriptor constraints

Let $f_m$ and $g_m$ be the descriptor functions corresponding to the point of the shapes $M$ and $N$ respectively, if the descriptor of the points is multi-scale, then $f_m(x) \in \mathbb{R}^k$ for every point $x$ indicates that it is a $k$-dimensional scale descriptor constraint. In this paper, we use HKS (Heat Kernel Signature) [21] as a descriptor constraint, because HKS inherits many nice properties from the heat kernel, such as being intrinsic and stable against perturbations of the shape. More remarkably, the set of HKS at all points on the shape fully characterizes the shape up to isometry.

Let $M$ be a compact Riemannian manifold, assuming that $x$ is a source point on $M$ at the time of $t=0$. According to the heat diffusion equation, there exists a function $h_t^M(x,y): \mathbb{R}^+ \times M \times M \to \mathbb{R}$ such that

$$H_t f(x) = \int_M h_t^M(x,y) f(y) dy \tag{7}$$

where $H_t f(x)$ denotes the heat diffusion distribution function at time $t$, and $h_t^M(x,y)$, called heat kernel, can be thought of as the amount of heat that is transferred from $x$ to $y$ at time $t$.

Given a point $x$ on the manifold $M$, its Heat Kernel Signature $\text{HKS}(x): \mathbb{R}^+ \to \mathbb{R}$ is defined as [21]:

$$\text{HKS}(x,t) = h_t(x,x) = \sum_{i=1}^{k} e^{-\lambda_i t} \phi_i^2(x) \tag{8}$$

where $\lambda_i$ and $\phi_i$ are the $i^{\text{th}}$ eigenvalue and the corresponding eigenfunction of the Laplace-Beltrami operator, respectively. We calculate the HKS for each model by uniformly sampling 100 points on a logarithmic scale over the time interval [$t_{min}$, $t_{max}$] with $t_{min}=4\ln 10/\lambda_{300}$, and $t_{max}=4\ln 10/\lambda_2$. A sparse Laplace matrix is obtained by using the mesh Laplace operator, and both the eigenvalues and eigenvectors are calculated to gain values of HKS of mesh models.

Now we represent the function $f_m$ as a vector of coefficients $\boldsymbol{a} = (a_0, a_1, \cdots a_i, \cdots)$ and $g_m$ as a vector $\boldsymbol{b} = (b_0, b_1, \cdots b_i, \cdots)$, then the formula (2) can be simply expressed as:

$$b_j = \sum_i a_i c_{ij} \tag{9}$$

where $c_{ij}$ is independent of $f_m$ and is completely determined by the bases and the map $T$. The matrix $\boldsymbol{C}$ is interpreted as a mesh registration of $\Phi^M$ and $\Phi^N$, that is $\boldsymbol{C}\Phi^M = \Phi^N$, here $\Phi^M$ and $\Phi^N$ are eigenfunctions for the shape $M$ and $N$, which indicates correspondence between the shape $M$ and $N$. Thus, the correspondence between shapes is transformed into a computed matrix $\boldsymbol{C}$ by functional maps.

## 5 Experimental Results

### 5.1 Data Sets

The experiments reported in the sequel were performed on the COSEG library [8] and Princeton Segmentation Benchmark library [22]. COSEG library contains candelabra, chair, four-legged animals, vases, guitar, high pin tumblers, etc. Note that, the sets of man-made shapes are composed of objects that possess significant variability, i.e., a common type of part can appear with different topologies and geometries across the set, and it can be absent or appear multiple times on a shape. For example in Figure 4, some candelabras have a base and a handle, while others only have a base, the geometry of their base is also varies significantly. The Princeton Benchmark library comprises a data set with 4,300 manually generated segmentations for 380 surface meshes of 19 different object categories.

5.2 **Results**

In our algorithm, we only need to set up a parameter $n$, the initial number of segmented components, the reasonable segmentation results will be produced. Because the proposed multi-scale feature descriptor is a good representation for various shapes and the functional maps can also construct correspondences of the same semantic components between shapes, we can naturally obtain good co-segmentation results with distinct segmented parts. Although the models in the container category and the candelabra category as shown in Figure 4 are topologically different, especially shapes in the container category containing different structural handles, or not, their same semantic components can be segmented using our unsupervised co-segmentation algorithm.

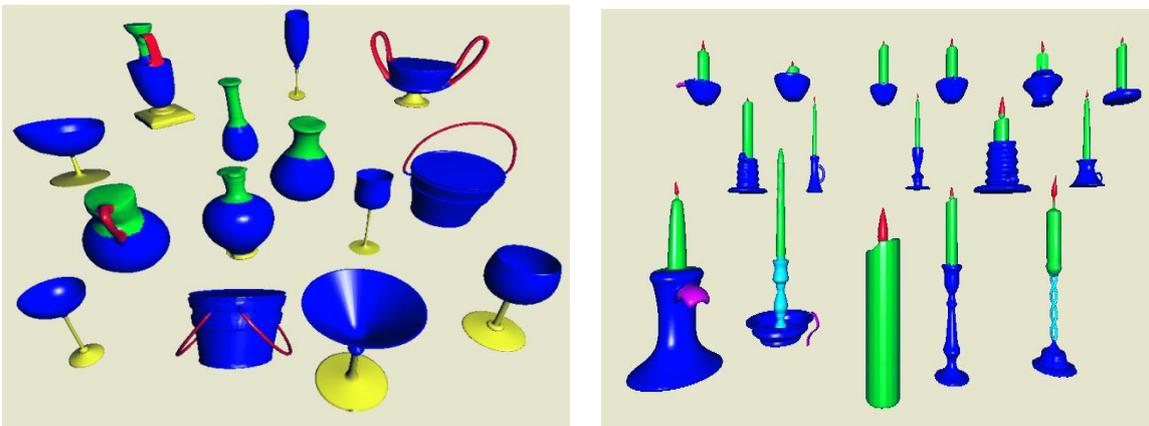

(a) Container              (b) Candelabra

Figure 4 Co-segmentation results of our algorithm

Similarly, when the proposed unsupervised co-segmentation algorithm is applied to biological shapes, such as the four-legged animals, birds as shown in Figure 5, it still could complete consistent segmentation. Thus, our algorithm shows good insensitivity to poses and shape variations.

Moreover, we use the co-segmentation results provided by the Princeton Segmentation Benchmark library as ground truth to evaluate robustness of the proposed algorithm. Our algorithm proves to be consistent in classifying all semantic parts as shown in Figure 1 and Figure 6.

We have compared our algorithm with the state-of-the-art descriptor-space spectral clustering(DSSC) method [6] as shown in Figure 7. The DSSC method returns the same semantic component for the headstock and fretboard of guitar, consequently resulting in improper segmentation results as shown in Figure 7(a), while our algorithm can establish a distinction between the instrument headstock, fretboard and body, thus it produces reasonable consistent segmentation results as shown in Figure 7(b). The DSSC method failed to partition the various segments of the guitars because the dissimilar parts may be linked through third-parties present in the set. The links are derived from the pairwise similarities between the parts' descriptor. Nevertheless, these problems appear due to imperfections in the clustering, which assign these parts with an incorrect label.

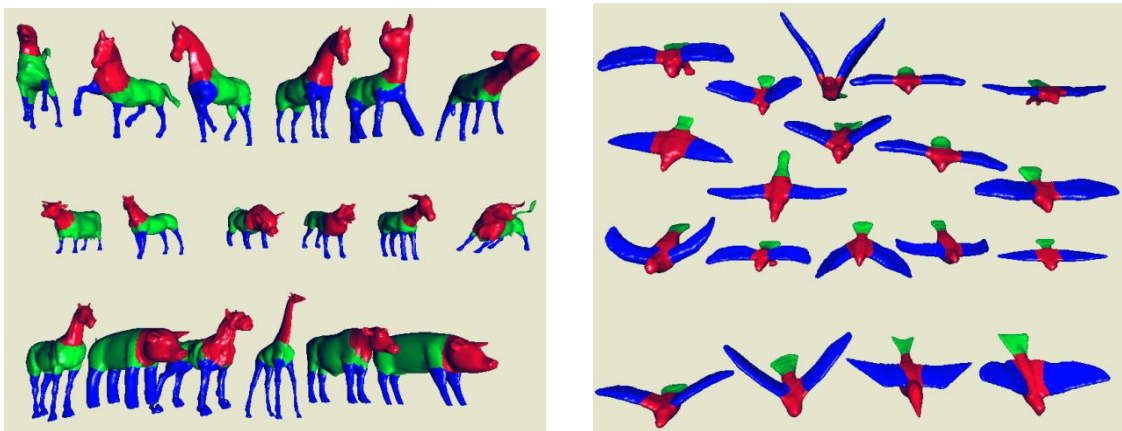

(a) Four-legged animals　　　　　　　　　　　　(b) Birds

Figure 5 Co-segmentation results of our algorithm

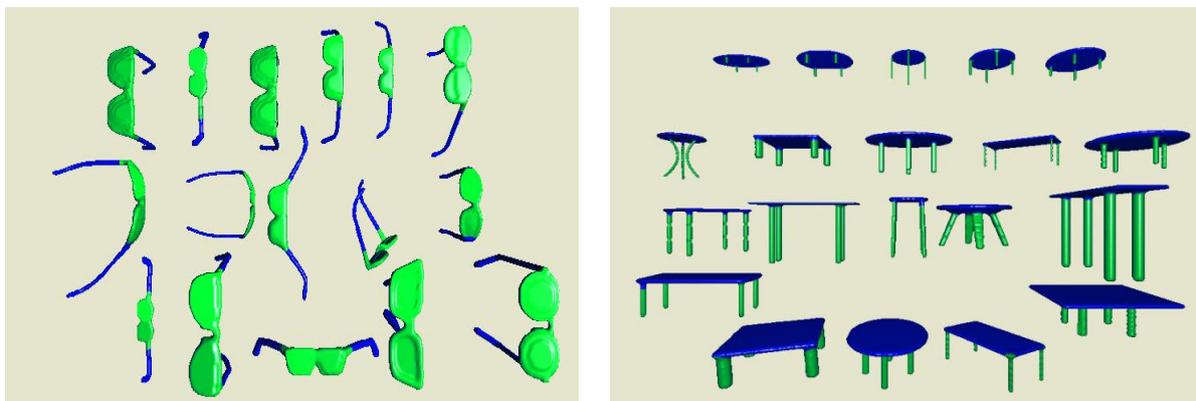

(a) Glasses　　　　　　　　　　　　(b) Table

Figure 6 Co-segmentation results of our algorithm

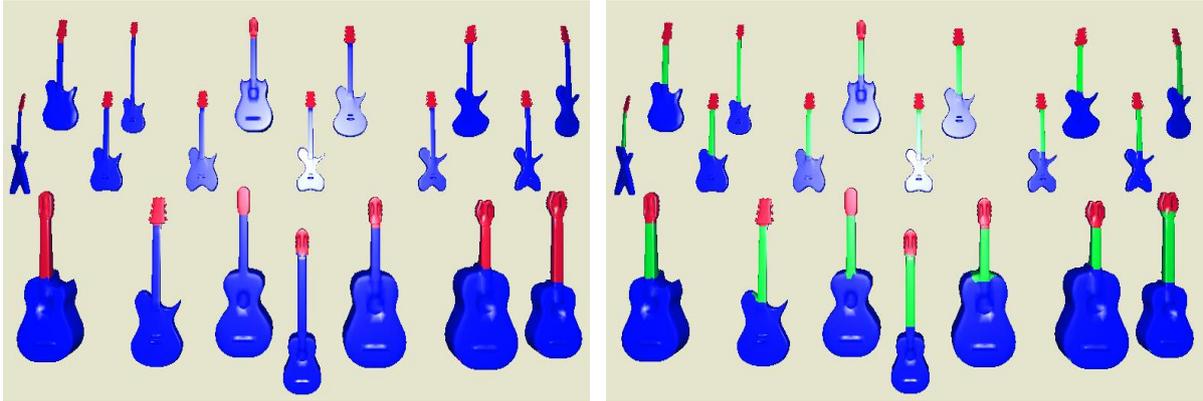

(a) Co-segmentation results of DSSC method [6]　　　(b) Co-segmentation results of our algorithm

Figure 7 Comparison of segmentation results of guitar set

Compared with subspace clustering(SC) method [7], the proposed algorithm produces superior results as shown in Figure 8. We can observe that the segmented parts of the plier clearly show overlap between the nose and handle by using subspace clustering method as shown in Figure 8(a), and both segments are classified as same. A very important reason may be due to the fact that the topological and geometrical composition of both the handle and nose of the plier are very similar, resulting in high similarity of the calculated affinity matrix, thus they are clustered as the same class. In this paper, we use the grid Laplace as the feature descriptor to get the most natural characteristic of 3D shapes, and this identifies segments with the geometric topology similarity but the different semantic information, as shown in Figure 8(b).

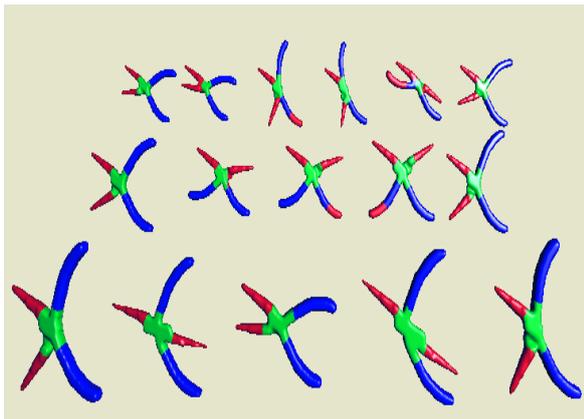 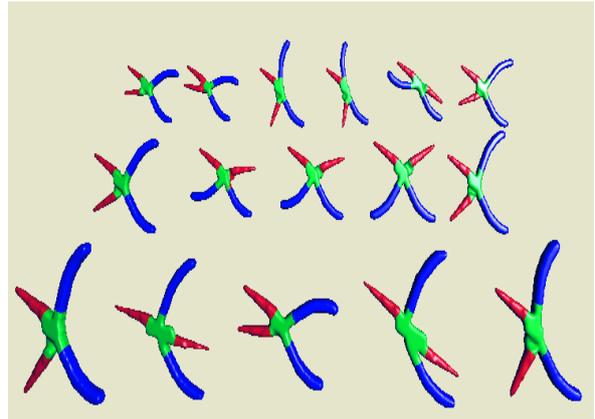

(a) Co-segmentation results of SC method [7]　　　(b) Co-segmentation results of our algorithm

Figure 8 Comparison of segmentation results of plier set

## 6 Conclusion

In this paper we present an unsupervised approach for consistently segmenting a set of shapes from a common category via functional map. The question inherent to co-analysis is whether we can extract more information by analyzing a set of shapes simultaneously, instead of individual or pairs of shapes. The key contribution lies in calculating an optimal correspondence between semantically similar parts through functional maps and labelling such parts. Experimental results demonstrate that the proposed algorithm can properly extract consistent parts across the shape set, achieving comparable accuracy to the state-of-the-art methods.

## Acknowledgments

The authors would like to thank the reviewers for their valuable comments. This work is supported by Natural Science Foundation of China under grant No. 61462059, China Postdoctoral Science Foundation Funded Project under grant No. 2013M542396, and Fundamental Research Funds for Gansu Provincial Universities under grant No. 214142.